\documentclass[12pt]{article}
\usepackage{amsmath,amssymb,amsthm,cite}
\usepackage[matrix,arrow]{xy}

\advance\topmargin by -\headheight
\advance\topmargin by -\headsep
\evensidemargin=\oddsidemargin
\renewcommand{\a}{\alpha}           
\newcommand{\al}{{\mathfrak{a}}}    
\renewcommand{\b}{\beta}            
\newcommand{\C}{\mathbb{C}}         
\newcommand{\Dl}{\Delta}            
\newcommand{\dl}{\delta}            
\newcommand{\E}{\mathfrak{E}}       
\newcommand{\eps}{\varepsilon}      
\newcommand{\Ga}{\Gamma}            
\newcommand{\g}{\mathfrak{g}}       
\newcommand{\ga}{\gamma}            
\newcommand{\id}{{\mathrm{id}}}     
\newcommand{\m}{\mu}                
\newcommand{\n}{\nu}                
\newcommand{\op}{\oplus}            
\newcommand{\ox}{\otimes}           
\renewcommand{\P}{\mathfrak{P}}     
\newcommand{\pa}{\partial}          
\newcommand{\R}{\mathbb{R}}         
\newcommand{\Rr}{\mathcal{R}}       
\newcommand{\sepword}[1]{\quad\mbox{#1}\quad} 
\newcommand{\set}[1]{\{\,#1\,\}}    
\newcommand{\Th}{\Theta}            
\newcommand{\thalf}{\tfrac{1}{2}}   
\renewcommand{\th}{\theta}          
\newcommand{\tihalf}{\tfrac{i}{2}} 
\newcommand{\U}{\mathcal{U}}        
\newcommand{\x}{\times}             
\renewcommand{\.}{\cdot}            
\let\hash=\#
\renewcommand{\#}{\mathbin{\hash}}  
\def\<#1,#2>{\langle#1,#2\rangle}   
\def\wick:#1:{\mathopen:#1\mathclose:} 

\makeatletter
\def\section{\@startsection{section}{1}{\z@}{-3.5ex plus -1ex minus
			  -.2ex}{2.3ex plus .2ex}{\large\bf}}
\def\subsection{\@startsection{subsection}{2}{\z@}{-3.25ex plus -1ex
			  minus -.2ex}{1.5ex plus .2ex}{\normalsize\bf}}
\renewcommand{\@dotsep}{200} 
\makeatother

\numberwithin{equation}{section}    

\theoremstyle{plain}

\theoremstyle{definition}

\theoremstyle{remark}

\begin{document}

\title{Hidden symmetry and Hopf algebra}




\vskip 35pt
\centerline{\Large\bf{Hidden symmetry and Hopf algebra}}

\vskip 35pt
\centerline{\Large J.~M. Gracia-Bond\'{\i}a${}^{a,b}$}

\vskip 10pt
\begin{center}

{\it ${}^a$ Departamento de F\'{\i}sica Te\'orica I, Universidad
Complutense\\ 28040 Madrid, Spain}\\
and\\
{\it ${}^b$ Departamento de F\'{\i}sica, Universidad de Costa Rica\\
2060 San Pedro, Costa Rica}

\end{center}

\vspace{30pt}

\rightline{\it In celebration of fecund}
\rightline{\it Pep\'{\i}n Cari\~nena's 60th birthday}

\vspace{30pt}

\centerline{\bf Abstract}
\medskip
{\leftskip=40pt \rightskip=40pt

We spell two conundrums, one of physical and another of mathematical
nature, and explain why one helps to elucidate the other. \par}

\vspace{25pt}




\section{Introduction I: Hopf algebra cohomology}
\label{sec:introibo}

Let us start by the mathematical conundrum. The two main classical
examples of Hopf algebras, respectively cocommutative and commutative,
are the enveloping algebra $\U(\g)$ of a Lie algebra~$\g$ and the
algebra~$\Rr(G)$ of representative functions on a group~$G$. For
definiteness, consider both over the complex numbers. On~$\U(\g)$ the
coproduct $\Dl:\U(\g)\to\U(\g)\ox\U(\g)$ is defined first on
elements~$X\in\g$ by
\begin{equation*}
\Dl(X) := X_{(1)} \ox X_{(2)} = X \ox 1 + 1 \ox X,
\end{equation*}
and then extended to all of~$\U(\g)$ multiplicatively. The output
of~$\Dl$ is invariant under exchange of the two copies of~$\U(\g)$ in
its image: this is cocommutativity. For the second, $\Rr(G)$ is the
space of functions $f:G\to\C$ whose translates $x\mapsto f(xt)$, for
all $t\in G$, generate a finite-dimensional subalgebra of the
commutative algebra of continuous functions~$C(G)$ under ordinary
multiplication. Then also $\Rr(G)$ is endowed with a coproduct in
which
\begin{equation*}
\Rr(G) \ox \Rr(G) \ni \Dl f \sepword{is given by} \Dl f(x,y) :=
\bigl(f_{(1)} \ox f_{(2)}\bigr)(x,y) := f(xy);
\end{equation*}
which is not cocommutative, unless~$G$ is abelian. There is of course
a functor going back from commutative Hopf algebras to groups.

When $\g$ is the tangent Lie algebra of a Lie group~$G$, it is
sometimes asserted that both previous constructions are mutually dual.
Reality is richer: although $\U(\g)$ is certainly in duality
to~$\Rr(G)$, there is a bigger dual space, the Sweedler dual
$\Rr^\circ(G)$ of~$\Rr(G)$, which is still a Hopf algebra, and
includes in particular the (also cocommutative) \textit{group algebra}
$\C G$; $\Dl g=g\ox g$ holds for `pure' elements~$g\in\C G$. In
fact, $\Rr^\circ(G)$ is a semidirect product of~$\C G$ and~$\U(\g)$.
We touch here at the general situation, as any cocommutative Hopf
algebra is a semidirect product of a group algebra and an enveloping
algebra~\cite{Murphyslaw1, Murphyslaw2}. For the general background on
Hopf algebras and matters of notation, besides~\cite{Murphyslaw2} we
refer to~\cite{Quaoar}; we denote by~1 the unit in~$H$ and the
augmentation homomorphism by~$\eta$.
Going to cohomology, is stands to reason that the cohomology of
enveloping algebras will contain the same information as the theory of
Lie algebra extensions, and that of commutative Hopf algebras as the
theory of group cocycles. But this is not quite what happens! Let us
follow Majid~\cite{Maje} now. With $\id$ the identity map of~$H$ onto
itself, define four maps from $H\ox H$ to~$H\ox H\ox H$ by:
\begin{equation*}
\Dl_0(\.) = 1\ox(\.); \quad \Dl_3(\.) = (\.)\ox1; \quad \Dl_1 =
\Dl\ox\id; \quad \Dl_2 = \id\ox\Dl.
\end{equation*}
Let $\chi$ be an invertible (in the algebra) element of~$H\ox H$. This
is a 2-cochain in general. Then its coboundary:
\begin{equation*}
H\ox H\ox H \ni \pa\chi :=
\Dl_0(\chi)\Dl_2(\chi)\Dl_1(\chi^{-1})
\Dl_3(\chi^{-1}) =: \pa_+\chi\pa_-\chi^{-1}.
\end{equation*}
An $2$-cocycle \textit{for}~$H$ is a $2$-cochain such that
$\pa\chi=1$. We compute:
\begin{align}
&(1\ox\chi)(\id\ox\Dl)\chi(\Dl\ox\id)\chi^{-1}(\chi^{-1}\ox1) = 1,
\nonumber \\
&\sepword{that is} (1\ox\chi)(\id\ox\Dl)\chi =
(\chi\ox1)(\Dl\ox\id)\chi.
\label{eq:the-rub}
\end{align}
Now, for~$H=\Rr(G)$, we recognize a group 2-cocycle, that is a nowhere
vanishing function~$\chi$ on~$G\x G$ such that
\begin{equation*}
\quad \chi(g_1,g_2)\chi(g_1g_2,g_3) =
\chi(g_1,g_2g_3)\chi(g_2,g_3); \quad \forall g,g_1,g_2,g_3 \in G.
\end{equation*}
We have recovered the standard theory of group 2-cocycles, allowing to
construct group extensions, and well known to physicists ---we require
as well unitality of~$\chi$, that is $(\eta\ox\id)\chi=
(\id\ox\eta)\chi=1$, guaranteeing $\chi(g,1_G)=\chi(1_G,g)=1$.

There naturally exists a dual theory of cocycles \textit{on} Hopf
algebras that, when applied to $\U(\g)$, reproduces the results of Lie
algebra cohomology. For Lie algebras like~$\P$, the one of the
Poincar\'e group, which is well known to be inextensible, such dual
procedure gives nothing; and of course the same is true of the
previous theory of cocycles \textit{for} Hopf algebras when applied to
the component of the identity of the Poincar\'e group. But, what about
coming back to the framework of~\eqref {eq:the-rub} and trying to
apply it to the \textit{noncommutative} Hopf algebra~$\U(\P)$? Or, for
that matter, what about trying to apply the theory of cocycles
\textit{on} Hopf algebras to~$\Rr(G)$? Well, it is not true that the
theory of~$n$-cocycles \textit{for} Hopf algebras, when used on
objects that are not commutative; or the dual theory of~$n$-cocycles
\textit{on} Hopf algebras, when used on non-cocommutative algebras,
always lead to proper cohomologies. But this was never to stop quantum
group theorists; and, lo and behold, for 2-cocycles there is no
difficulty, indicating that a sort of generalized symmetry is present.
A 2-cocycle \textit{for}~$\U(\g)$ is precisely what they call a
\textit{twist}. We briefly review how twists permit to deform the
coproduct. Let~$H$ be a cocommutative Hopf algebra and~$\chi$ the
twist. Consider $\Dl_\chi(a) := \chi\Dl(a)\chi^{-1}$. This gives a new
coproduct on~$H$. First of all, for the new coproduct $\Dl_\chi$ is
still an algebra map:
\begin{equation*}
\Dl_\chi(ab) = \chi\Dl(ab)\chi^{-1} = \chi\Dl(a)\Dl(b)\chi^{-1} =
\chi\Dl(a)\chi^{-1}\chi\Dl(b)\chi^{-1} = \Dl_\chi(a)\Dl_\chi(b).
\end{equation*}
Let us check coassociativity of~$\Dl_\chi$:
\begin{align*}
(\Dl_\chi\ox\id)\Dl_\chi(a) &=
\chi_{12}(\Dl\ox\id)(\chi\Dl(a)\chi^{-1})\chi^{-1}_{12}
\\
&= \chi_{12}((\Dl\ox\id)\chi)((\Dl\ox\id)\Dl
a)((\Dl\ox\id)\chi^{-1})\chi^{-1}_{12}
\\
&= \chi_{23}((\id\ox\Dl)\chi)((\id\ox\Dl)\Dl
a)((\id\ox\Dl)\chi^{-1})\chi^{-1}_{23} = (\id\ox\Dl_\chi)\Dl_\chi(a).
\end{align*}
Here $\chi_{12}$ of course means $\chi\ox1\in H\ox H\ox H$, and so on.
We have used~\eqref{eq:the-rub}. The resulting Hopf algebra is
denoted~$H_\chi$. Naturally this twisting procedure to create new Hopf
algebras, when used with cohomologous cocycles, gives Hopf algebras
that are isomorphic via inner automorphisms; but often an appropriate
twist gives a novel construction.

Now, for $H$ any Hopf algebra, not necessarily commutative or
cocommutative, a left (Hopf) $H$-module algebra $(A,\x)$ is a not
necessarily commutative algebra which is a representation space for
(the algebra structure of)~$H$, and moreover
\begin{equation*}
h\.(a\x b) = h_{(1)}\.a\x h_{(2)}\.b
\sepword{whenever}  h \in H,\  a,b \in A.
\end{equation*}
The formula $h\.1_A=\eta(h)1_A$ usually added here is
redundant~\cite{Miriam}. Write also $\x(a\ox b)=a\x b$. The
consequence of the twist is that the product
\begin{equation}
a\star_\chi b := \x\bigl(\chi^{-1}(a\ox b)\bigr),
\label{eq:stilted-form}
\end{equation}
for $a,b\in A$, defines a new associative algebra~$A_\chi$, covariant
under~$H_\chi$. In effect, associativity of~$\star_\chi$ follows from
the 2-cocycle condition. One trivially checks covariance: for $h\in
H$,
\begin{align*}
h\.(a\star_\chi b) &:= h\.\x\bigl(\chi^{-1}(a\ox b)\bigr) =
\x\bigl(\Dl(h)\.\chi^{-1}(a\ox b)\bigr)
\\
&= \x\bigl(\chi^{-1}\Dl_\chi(h)\.(a\ox b)\bigr) =:
\star_\chi\bigl(\Dl_\chi(h)\.(a\ox b)\bigr).
\end{align*}
We ask forgiveness from the reader for the heavy notation; it will be
needed later.

\medskip

The mystery is this: in principle there is \textit{no more} information
in the Hopf algebra $\U(\g)$ than in the Lie algebra~$\g$. So, in terms
of symmetry, what may the twisting procedure mean?

\medskip

For~$\P$ with its usual generators $T_\nu,M_{\a\b}$, taking
$\chi_\Th:= \exp(-\tihalf\th^{\m\n}\, T_\mu \ox T_\nu)$, where
$\Th:=(\th^{\m\n})$ is a skew-symmetric matrix, a little calculation
with a glance at~\eqref{eq:ciao-ciao-bambina} gives:
\begin{equation*}
\Dl_\chi(M_{\a\b}) = M_{\a\b} \ox 1 + 1 \ox M_{\a\b}
+ \tihalf\th^{\rho\sigma}\bigl((g_{\a\rho}T_\b - g_{\b\rho}T_\a) \ox
T_\sigma + T_\rho \ox (g_{\a\sigma}T_\b -
g_{\b\sigma}T_\a)\bigr);
\end{equation*}
while the coproduct for the~$T_\nu$ is not modified. This is, in
cohomological terms, what was done in~\cite{HeartOfDarkness}. As in
that reference, to check the cocycle condition~\eqref{eq:the-rub} is
left to the reader. Moreover, $\star_{\chi_\Th}$ is a Moyal
product~\cite{Moyal}, and the apparently unlikely deed of having the
Poincar\'e Lie algebra act on Moyal algebra has been done. Deforming
the coproduct of an enveloping algebra is much less drastic than
deforming the product, and, very gratifyingly, the Casimirs and the
whole paraphernalia of relativistic fields remain unaffected. Still,
the manner in which the action of~$\P$ on its representation spaces
propagates to their tensor products has been modified.

\section{Introduction II: a physical discussion}
\label{sec:introibobis}

The motivation for~\cite{HeartOfDarkness} was that the question of
relativistic symmetry on noncommutative Minkowski (or Euclidean, as
the case may be) spacetime is apparently a vexing one. Indeed,
periodically, and as recently as~\cite{OldCudmurgeon}, there are
complaints about the calamitous state of the study of covariance in
noncommutative field theory (NCFT). Often, authors just look at the
Moyal commutators $[x^\mu,x^\nu]_{\star_\Th}=i\th^{\mu\nu}$ and
conclude that Poincar\'e invariance is broken down to a subgroup. To
wit, with $\P_n,\E_n$ respectively denoting the $\thalf(n^2+n)$
dimensional Poincar\'e (respectively, Euclidean) Lie algebras of
Lorentz transformations (respectively, rotations) and translations
on~$\R^n$, and assuming~$\Th$ has maximal rank, $\P\equiv\P_4$ reduces
to a six-dimensional Lie algebra isomorphic to~$\P_2\op\E_2$.
See~\cite{LAGMAVAZ}. Respectively, $\E_4$ would break down
to~$\E_2\op\E_2$. One easily finds in the NCFT literature statements
like: ``the physics depends on the frame of
reference''~\cite{Deprimente}; picturesquely adding that it must be
so, because the speed of light in a noncommutative geometry depends on
the direction of motion. Also~\cite{Lamentable} espouses the viewpoint
of~\cite{Deprimente}. On the face of it, this is a defensible one.

\medskip

But if so, would it be justified to use the Wigner particle structure
of scalar, vector, spinor fields and so on, as done as a matter of
course in almost every paper in~NCFT? This is the second mystery.

\medskip

This is why the ideas in paper~\cite{HeartOfDarkness} ---see
also~\cite{Wesstercia}--- were welcome. These authors apparently
establish that a form of Poincar\'e covariance is relevant in NCFT. In
particular, Poincar\'e group representations and their tensor products
are totally pertinent. Later, it has been claimed that the analysis
of~\cite{HeartOfDarkness} extends to the conformal
group~\cite{Calcolini}; also twisted conformal symmetry in NCFT in two
dimensions has been examined~\cite{PassageToIndia}.

There is, however, a touch of obscurantism in~\cite{HeartOfDarkness}.
For a start, their treatment is couched in the abstract language of
quantum groups, and no physical interpretation of their reconstruction
of Moyal algebra from a twisting of the coproduct on the Poincar\'e
enveloping algebra was attempted. Also, the twisting is a general
geometrical fact, not specifically linked to the the Poincar\'e group.

Actually, some of the early treatments of relativistic symmetry in
NCFT are more forthcoming: we refer to the lucid remarks
in~\cite{ChristmasCarroll} and the analysis in the deep
paper~\cite{CircleOfVienna}. One may rephrase their argument as
follows. Assume that in a region of the space there is a background
field. Its presence modifies the vacuum, breaking Poincar\'e
invariance, in the sense that active Lorentz transformations are no
longer symmetries of the physical system. But this does not stop an
electron under the influence of that background from being a
relativistic electron. It remains possible for observers to describe
the system in a Poincar\'e covariant way, by suitable changes in the
description of the background (the so-called observer or passive
Lorentz transformations). In doing so, one stays within the same
theory; were we to modify charges, masses or other internal variables
of the system it would be otherwise. The phenomenon of variation of
the speed of light is expected with any background field~\cite{LPR}; a
result that does not contradict relativity: rather, relativity is used
to derive it.

Now, one may contend that the situation is analogous in NCFT, where
one has the skew-symmetric tensor~$\Th$ describing the background.
Noncommutativity would brook no ether, even we ignore as yet what its
dynamical equations ---and boundary conditions--- are. The origin of
the theory in string dynamics~\cite{SeibergW} does not appear to
contradict this view. Our analysis has points of contact with the
recent papers~\cite{PolishedPolish}, that use the Hopf dual
$H^\circ_\chi$ of $H_\chi$, and with~\cite{BanerjeeChK}. One should
rather not speak alternatively of unbroken/broken symmetry, but of
\textit{manifest/hidden} invariance. Relativistic symmetry is simply
hidden in NCFT.

The solution of the second conundrum holds a key to the first.
Symmetries in the noncommutative regime (no less than in the
commutative one) are \textit{always} described by automorphisms
---that is, derivations at the infinitesimal level--- of the algebra
of observables. When the symmetry is hidden, those derivations involve
the parameters of the vacuum state. We discover that, in some cases at
least, twists or deformations of Hopf algebras are related to hidden
symmetries.

\section{Conventions}
\label{sec:so-so}

The form of the Moyal product used in this paper is that of
Rieffel~\cite{RieffelDefQ}; this is good for any rank of~$\Th$, and is
moreover an exact (nonperturbative in~$\Th$) deformation of the
ordinary product. Due to the singular nature of the $\Th\downarrow0$
limits, all kind of pitfalls await the unwary user of perturbative
forms. For the precise relation between both kinds we refer
to~\cite{Nereid}, in the analogous framework of phase space Quantum
Mechanics. Of course, at some points we need to fall back on
perturbative forms for comparison purposes. Given the $4\x4$
skewsymmetric matrix $\Th$, the Moyal star product on~$\R^4$ is:
\begin{equation}
f \star_\Th h(x) = \frac{1}{(2\pi)^4} \int d^4\!y \,d^4\!u\,f\bigl(x +
\thalf\Th y\bigr)\,h(x + u)\,e^{iy\. u}.
\label{eq:Moyal-prod}
\end{equation}

The group $A(4;\R)$ of affine transformations acts on four-vectors
by~$x\mapsto Lx + a$, where $a\in\R^4$ and $L$ denotes a matrix with
$\det L\ne0$. We have $(L,a)(L',a') = (LL',La' + a)$. Thus the inverse
transformation of~$(L,a)$ is $(L^{-1},-L^{-1}a)$. Often we write
just~$g$ for~$(L,a)$ and $g\.x$ for $Lx + a$. An action on functions
on~$\R^4$ ensues, of the form:
\begin{equation*}
[(L,a)\triangleright f](x) := f\bigl(L^{-1}(x - a)\bigr).
\end{equation*}
This definition leads to the natural $g_1\triangleright
[g_2\triangleright f]=(g_1g_2)\triangleright f$.

The 11-dimensional Weyl group~$W$ of \textit{rigid} conformal
transformations (translations and dilations plus Lorentz
transformations) generated by~$\set{T_\tau,D,M_{\a\b}}$, with
commutation relations:
\begin{align}
[T_\tau, T_\sigma] &= 0; \qquad [T_{\tau}, D] = T_{\tau}; \qquad
[T_\tau, M_{\a\b}] = g_{\tau\a}T_\b - g_{\tau\b}T_\a;
\nonumber \\
[ D, M_{\a\b}] &= 0; \quad [M_{\a\b}, M_{\ga\dl}] = g_{\b\ga}M_{\a\dl}
+ g_{\a\dl}M_{\b\ga} - g_{\a\ga}M_{\b\dl} - g_{\b\dl}M_{\a\ga},
\label{eq:ciao-ciao-bambina}
\end{align}
will be envisaged. This subgroup of~$A(4;\R)$ is singled out in
relation with dynamical ---as opposed to merely geometrical---
aspects: for definiteness we consider now $\star$-gauge
(noncommutative Yang--Mills) theories, whose action is invariant
under~$W$. This is as in~\cite{CircleOfVienna}. The prototype is the
Maxwell-$\star$ theory on~$\R^4$, with gauge potential $A_\m$.
Unfortunately, lack of space prevents us from going into the
particulars of gauge theory: almost solely its vector aspect is
important here. Throughout, we consider~$\R^4_\Th$ with constant
(position-independent) noncommutativity. Let us note, however, that
the interplay between coordinate, gauge and~$\Th$-variables
characteristic of NCFT is even more patent in non-constant
noncommutativity spaces~\cite{Melpomene}, of whose the one considered
here must be regarded as a limit case.

\section{Twisted affine transformations}
\label{sec:going-haywire}

The question is to compute $[g\triangleright f]\star_\Th
[g\triangleright h]$. Denote by~$L^{-t}$ the contragredient matrix
of~$L$. By a simple change of variables in the
integral~\eqref{eq:Moyal-prod} one obtains:
\begin{align}
&[g\triangleright f] \star_\Th [g\triangleright h](g\.x) = 
f \star_{L^{-1}\Th L^{-t}} h(x);
\sepword{that is to say,}
\nonumber \\
& [g\triangleright f] \star_\Th [g\triangleright h] =
g\triangleright(f \star_{L\Th L^t} h).
\label{eq:crux-of-the-matter}
\end{align}
In the noncommutative world, i.e., for $\Th\ne0$, spacetime and parameter
transformations are intimately linked; we see
in~\eqref{eq:crux-of-the-matter} emerging an action, trivial for
translations, of the affine group on the linear space of
skewsymmetric matrices, given by
\begin{equation*}
(L,a)\.\Th = L\.\Th := L\Th L^t.
\end{equation*}
There is neither novelty nor mystery about this action: it is just
classical congruence, studied by Lagrange and Sylvester centuries ago.
Its only invariant is the rank, so the orbits are constituted
respectively by the generic set of invertible skewsymmetric
matrices, the set of non-invertible, nonvanishing skewsymmetric
matrices, and the zero matrix. Given~$\Th$, the matrices $L\in
A(4;\R)$ such that $L\.\Th=\Th$ form a ``little group'' $A_\Th$, of
dimension~10 for the generic orbit (then and only then does $\Th$
define a symplectic form). There is of course an enormous difference
between merely regarding $A_\Th$ ---or $A_\Th\cap W$--- as `the'
symmetry group, and regarding it as the result of a symmetry breaking
$A(4;\R)\downarrow A_\Th$ of a larger group.

In summary, on the variables~$(x,\Th)$ the affine transformations
act by
\begin{equation}
(L,a)\.(x,\Th) = (Lx + a, L\Th L^t).
\label{eq:tapa-del-perol}
\end{equation}
For the induced action on the sections of the field of
$\star$-algebras over the space of all~$\Th$'s, regarded as functions
of~$(x,\Th)$, from~\eqref{eq:crux-of-the-matter} we conclude that
\begin{equation}
[g\triangleright f] \star_\Th [g\triangleright h] = g\triangleright
(f\star_\Th h).
\label{eq:mama-de-Tarzan}
\end{equation}
Such an automorphism equation is the trademark of covariance. The
paper is but a corollary of this fundamental formula. Incidentally,
the oldest avatar of these formulae we know of was found
in~\cite{Formulas}. Also, recently~\eqref{eq:mama-de-Tarzan} has been
rederived from a different viewpoint in~\cite{SantaCompanya}.

If $g\in A_\Th$, its action is vertical on that field, and then we may
replace~\eqref{eq:mama-de-Tarzan} by:
\begin{equation*}
[g\triangleright f] \star_\Th [g\triangleright h](x) = 
f\star_\Th h(g^{-1}\.x).
\end{equation*}
Moreover this equivariance can be realized by global gauge
transformations, that is, by conjugation with $\star_\Th$-unitary
elements. Properties of those unitaries were reported
in~\cite{Selene}.

\medskip

Next we descend to the infinitesimal level. The
action~\eqref{eq:tapa-del-perol} possesses infinitesimal generators,
which are vector fields in the $(x,\Th)$ spaces. As convenient
coordinates on the noncommutativity parameter sector we may take the
six nonvanishing components of~$\Th$. In some sense, this is whole
point: the variable is~$\Th$, the coordinates do not have intrinsic
physical meaning. Writing $L=1+B$ in~\eqref{eq:tapa-del-perol}, for
small~$B$ we have
\begin{equation*}
(L,a)\.(x,\Th) \sim (x + Bx + a,\Th + B\Th + \Th B^t) = \bigl(x + Bx +
a,\Th + B\Th - (B\Th)^t\bigr).
\end{equation*}
This means that suitable generators are
\begin{equation*}
\Ga_{B,a} := \bigl(a^\a + b^\a_\b x^\b\bigr)\frac{\pa}{\pa x^\a} +
\bigl(b^\rho_\b\th^{\b\sigma} + \th^{\rho\b}b^\sigma_\b\bigl)
\frac{\pa}{\pa\th^{\rho\sigma}} = \bigl(a^\a + b^\a_\b
x^\b\bigr)\frac{\pa}{\pa x^\a} +
(B\Th)^{[\rho\sigma]}\frac{\pa}{\pa\th^{\rho\sigma}},
\end{equation*}
where we have put $(b^\a_\b)=B$. We write $\pa_\a\equiv\pa/\pa x^\a$
and for a while omit from our considerations the $a^\a\pa_\a$ part: it
is well known that the Leibniz rule for these operators with the Moyal
product holds. The remaining vector fields~$\Ga_B$ have components
linear in the respective coordinates. We rewrite
\begin{equation}
\Ga_B = b^\a_\b x^\b\pa_\a + \bigl(b^\rho_\b\th^{\b\sigma} +
\th^{\rho\b}b^\sigma_\b\bigr)\frac{\pa}{\pa\th^{\rho\sigma}} =:
\eps^\a_B(x)\pa_\a -
\dl_{\eps_B}\th^{\rho\sigma}\frac{\pa}{\pa\th^{\rho\sigma}}.
\label{eq:now-what}
\end{equation}
The last form of the second part of~$\Ga_B$ points to its geometrical
meaning: it is (minus) the \textit{Lie derivative} with respect to the
vector field~$\eps_B$ of the contravariant components of the matrix
$\Th$, regarded as a tensor:
\begin{equation*}
\dl_\eps\th^{\rho\sigma} = \eps^\a_B(x)\pa_\a\th^{\rho\sigma} -
\th^{\b\sigma}\pa_\b\eps_B^\rho - \th^{\rho\b}\pa_\b\eps_B^\sigma =
-b_\b^\rho\th^{\b\sigma} - \th^{\rho\b}b_\b^\sigma =
-\dl_{\eps_B}\th^{\sigma\rho}.
\end{equation*}
This is an indication that we are on the right track. It is
obviously important ---in physics in relation with application of
Noether's theorem, for instance--- to record the $4\x4$ matrices~$B$
such that~$B\Th+\Th B^t=0$ or~$\dl_{\eps_B}\th^{\mu\nu}=0$. We
identify the Lie algebra~$\al_\Th$ of matrices~$B$ such that~$B\Th$ is
symmetric. Now, from~\eqref{eq:mama-de-Tarzan} we quote its
infinitesimal version
\begin{equation}
\Ga_B(f \star_\Th h) = \Ga_B f \star_\Th h + f \star_\Th \Ga_B h.
\label{eq:dangerous-turn}
\end{equation}
The simplicity of~\eqref{eq:dangerous-turn} \textit{and} of the path
leading to it is remarkable. For $B\in\al_\Th$, in view of our remark
at the end of the previous section, $\Ga_B$ is an inner derivation of
the $\star$-algebra (precisely, it is equivalent to a
$\star_\Th$-commutator in a multiplier $\star$-algebra)
for~$\det\Th\ne0$; otherwise $\Ga_B$ is outer. This kind of
derivations were not considered in the previous
analysis~\cite{LizzSZZ...}. In the simpler case of~$\R^2$, with
$b^\a_\b=\dl^\a_\b$ and $\th^{\a\b}=\eps^{\a\b}\th$, we get only the
derivation $x^\m\frac{\pa}{\pa x^\m}+2\th\frac{\pa}{\pa\th}$. This had
been noticed by some mathematicians~\cite{GKingJohn}.

The reader is encouraged to check~\eqref{eq:dangerous-turn} by
brute-force calculations: compute $\frac{\pa}{\pa\th^{\rho\sigma}}(f
\star_\Th h)$ and $\eps_B(f \star_\Th h)$, directly
from~\eqref{eq:Moyal-prod} in both cases, using integration by parts.

Summarizing: for~$\Th=0$ (the commutative world), automorphisms of the
algebra of observables are diffeomorphisms. These are locally
generated by vector fields, with components which are arbitrary in
principle. In the noncommutative world, vector fields no longer
represent infinitesimal symmetries. However, vector fields with
components up to degree one in the coordinates can still be
interpreted as ---manifest or hidden--- symmetries of Moyal algebra.

\section{Coming back to~\cite{CircleOfVienna}}
\label{sec:the-stress}

The comparison with~\cite{CircleOfVienna} is very instructive. All the
generators in~\eqref{eq:ciao-ciao-bambina} are affine.
In~\cite{CircleOfVienna} their action is written down only on the
(unquantized) gauge potentials and the gauge field strengths
$F_{\mu\nu}:=\pa_\m A_\n-\pa_\n A_\m-i[A_\m, A_\n]_{\star_\Th}$, in
terms of functional derivatives with respect to the former. That
method is unnecessarily complicated. It is enough to treat the
$A_\nu,F_{\mu\nu}$ as covariant vectors and 2-tensors, respectively,
and substitute the Lie derivative for the action~\eqref{eq:now-what}
of~$\Ga_B$ on scalar functions, for the corresponding matrix~$B$. For
the gauge potentials, as the $\pa/\pa\th^{\rho\sigma}$ do not
intervene, this gives
\begin{equation*}
\Ga_B(A_\mu) =  b^\rho_\tau x^\tau\pa_\rho A_\mu +
A_\rho\pa_\mu(b^\rho_\tau
x^\tau) = b^\rho_\tau x^\tau\pa_\rho A_\mu + b^\rho_\mu A_\rho.
\end{equation*}
In particular, if $B=M_{\a\b}$ then $b^\rho_\tau=\dl^\rho_\b
g_{\a\tau}-\dl^\rho_\a g_{\b\tau}$, and if $B=D$ then
$b^\rho_\tau=\dl^\rho_\tau$, so we get
\begin{equation*}
\Ga_{M_{\a\b}}(A_\mu) = x_\a\pa_\b A_\mu - x_\b\pa_\a A_\mu +
g_{\mu\a}A_\b - g_{\mu\b}A_\a; \quad \Ga_D(A_\mu) = x^\rho\pa_\rho
A_\mu + A_\mu;
\end{equation*}
together with $\Ga_{T_\tau}(A_\mu)=\pa_\tau A_\mu$, of course. For the
field strengths, one has to take into account the $\Th$-dependence in
their definition. Still the corresponding terms cancel and one
concludes
\begin{align*}
\Ga_{M_{\a\b}}(F_{\mu\nu}) &= x_\a\pa_\b F_{\mu\nu} - x_\b\pa_\a F_{\mu\nu}
+ g_{\mu\a}F_{\b\nu} - g_{\mu\b}F_{\a\nu} + g_{\nu\a}F_{\b\mu} -
g_{\nu\b}F_{\a\mu};
\\
\Ga_D(F_{\mu\nu}) &= x^\a\pa_\a F_{\mu\nu} + 2F_{\mu\nu}; \qquad
\Ga_{T_\tau}(F_{\mu\nu})=\pa_\tau F_{\mu\nu}.
\end{align*}
We have recovered in all simplicity the results
of~\cite{CircleOfVienna}, with the proviso that the widespread use of
$\star$-anticommutators in that reference is another unnecessary
complication, because $\thalf(x_\a\star_\Th\pa_\b F_{\mu\nu}+\pa_\b
F_{\mu\nu}\star_\Th x_\a)$ is the same as $x_\a\pa_\b F_{\mu\nu}$ for
any~$\Th$. All looks like in the commutative world, and invariance of 
the noncommutative Yang--Mills action ensues.

\section{Coming back to~\cite{HeartOfDarkness} and~\cite{Calcolini}}
\label{sec:the-stretch}

In this last discussion, our \textit{point de d\'epart}
is~\eqref{eq:stilted-form} for $\chi=\chi_\Th$. We write $\star_\Th$
for $\star_{\chi_\Th}$, giving the asymptotic version of the Moyal
product~\cite{SeibergW}, and $\Dl_\Th$ for $\Dl_{\chi_\Th}$. Now, let
$X$ be \textit{any} derivation of the commutative product~$\x$, i.e.,
any vector field. It has the property that
$$
X\.\x(a\ox b) = \x\bigl(\Dl_0(X)\.(a\ox b)\bigr).
$$
Then
\begin{align*}
X\.(a\star_\Th b) &= X\.\x\bigl(\chi^{-1}_\Th(a\ox b)\bigr) =
\x\bigl(\Dl_0(X)\.\chi^{-1}_\Th(a\ox b)\bigr)
\\
&= \x\bigl(\chi^{-1}_\Th\Dl_\Th(X)\.(a\ox b)\bigr) =
\star_\Th\bigl(\Dl_\Th(X)\.(a\ox b)\bigr).
\end{align*}
This is a general geometrical fact, independent of whether $X$ is the
generator of a Poincar\'e transformation or not. It is then scarcely
surprising that Matlock~\cite{Calcolini} has found it to be valid for
local conformal transformations. For similar reasons, sections~3
of~\cite{HeartOfDarkness} and 4 of~\cite{Calcolini} are tautological.

Next we need a more explicit name, say $\rho$, for the representation
of~$X$ as a Moyal algebra operator. What we have been able to prove in
the above is that, for~$X$ an \textit{affine} transformation, if
$\rho(\Dl_\Th (X))=\rho(\Dl_0(X))+R(X)$, then there is another linear
operation $\tilde\rho$ of~$X$ on the Moyal algebra, not a derivation
either, such that
$$
\tilde\rho(X)\.(a\star_\Th b) = \tilde\rho(X)\.a \star_\Th b +
a \star_\Th \tilde\rho(X)b - \star_\Th\bigl(R(X)(a\ox b)\bigr);
$$
and so $\rho+\tilde\rho$ is a $\star$-derivation. Thus hidden and
twist covariance boil respectively down to
\begin{equation*}
X^\Th m_\Th = m_\Th\Dl_0(X)
\qquad
\sepword{and} Xm_\Th = m_\Th\Dl_\Th(X),
\end{equation*}
where we have written $X^\Th$ for the realization of~$X$ as a
derivation in $(x,\Th)$-space. This does not seem to work for special
conformal transformations, as noted
in~\cite{CircleOfVienna,PolishedPolish,Vega}.

\section{Conclusion}
\label{sec:Zulu}

We have examined in parallel references~\cite{HeartOfDarkness}
and~\cite{CircleOfVienna}. This in particular amounts to a (partial)
physical interpretation of the manipulation in~\cite{HeartOfDarkness},
in terms of an action of the Poincar\'e group by observer
transformations, involving the space of parameters describing a
noncommutativity background field. For Euclidean symmetry everything
would work out essentially the same.

Our results show by the way that the seminal `quantum spacetime'
formalism by Doplicher, Fredenhagen and Roberts~\cite{Overwrought} and
NCFT as currently practiced essentially coincide. Mathematically, the
space of states in~\cite{Overwrought} is a particular orbit of
congruence~\eqref{eq:tapa-del-perol} when~$L$ is in the Lorentz group.
This should have been clear at least since reference~\cite{MarcWise}.
In the quantum spacetime formalism questions of relativistic symmetry
breaking can be adjourned for a while by use of the Heisenberg picture
for fields depending on the position variables; however, to perform
physical evaluations, one is forced to choose a state, that is, a
finite measure on the~$\Th$-space; and in so doing Lorentz symmetry
becomes hidden.

The moral of our story is that sometimes concrete group actions are
able to complement what `twisted symmetry' teaches us. It would be
good to know under which general conditions cocycles for cocommutative
Hopf algebras relate to hidden symmetry.

\subsection*{Acknowledgment}

The author thanks his collaborators in~\cite{Vega} ---which deals at
more length with the same subject--- and P.~Aschieri, D.~Bahns,
M.~Dubois-Violette and H. Grosse for discussions and suggestions. He
has been supported by MEC-Spain through grant~FIS2005-02309.


\begin{thebibliography}{99}
	
\bibitem{Murphyslaw1}
B. Kostant, unpublished.

\bibitem{Murphyslaw2}
M. E. Sweedler,
\textit{Hopf algebras},
Benjamin, New York, 1969.

\bibitem{Quaoar}
H. Figueroa and J. M. Gracia-Bond\'{\i}a,
Reviews in Math. Phys. {\bf 17} (2005)~881.

\bibitem{Maje}
S. Majid,
\textit{Foundations of Quantum Group Theory},
Cambridge University Press, Cambridge, 1995.

\bibitem{Miriam}
M. Cohen,
Contemp. Math. {\bf 134} (1992)~1.
						   
\bibitem{HeartOfDarkness}
M. Chaichian, P. P. Kulish, K. Nishijima and A. Tureanu,
Phys. Lett. B {\bf 604} (2004)~98.

\bibitem{Moyal}
J. E. Moyal,
Proc. Camb. Philos. Soc. {\bf 45} (1949)~99.

\bibitem{OldCudmurgeon}
B. Schroer,
Ann. Phys. {\bf 319} (2005)~92.
 
\bibitem{LAGMAVAZ}
L. Alvarez-Gaum\'e and M. A. V\'azquez-Mozo,
Nucl. Phys. B {\bf 668} (2003)~293.

\bibitem{Deprimente}
A. Iorio and T. S\'ykora,
Int. J. Modern Phys. A {\bf17} (2002)~2369.
								 
\bibitem{Lamentable}
R. Jackiw and S.-Y. Pi,
Phys. Rev. Lett. {\bf88} (2002)~111603.

\bibitem{Wesstercia}
J. Wess,
``Deformed coordinate spaces derivatives'',
hep-th/0408080.

\bibitem{Calcolini}
P. Matlock,
Phys. Rev. D {\bf71} (2005)~126007.

\bibitem{PassageToIndia}
F. Lizzi, S. Vaidya and P. Vitale,
``Infinite conformal symmetry in noncommutative two-dimensional
quantum field theory'',
hep-th/0601056.

\bibitem{ChristmasCarroll}
S. M. Carroll, J. A. Harvey, V. A. Kosteleck\'y, C. D. Lane and T.
Okamoto,
Phys. Rev. Lett. {\bf 87} (2001) 141601.

\bibitem{CircleOfVienna}
A. A. Bichl, J. M. Grimstrup, H. Grosse, E. Kraus, L. Popp, M. Schweda
and R. Wulkenhaar,
Eur. Phys. J. C {\bf24} (2002)~165.

\bibitem{LPR}
J. I. Latorre, P. Pascual and R. Tarrasch,
Nucl. Phys. B {\bf437} (1995)~60.
 
\bibitem{SeibergW}
N. Seiberg and E. Witten,
J. High Energy Phys. {\bf 9909} (1999)~032.

\bibitem{PolishedPolish}
C. Gonera, P. Kosi\'nski, P. Ma\'slanka and S. Giller,
Phys. Lett. B {\bf 622} (2005)~192;
Phys. Rev. D {\bf 72} (2005) 067702.

\bibitem{BanerjeeChK}
R. Banerjee, B. Chakraborty and K. Kumar,
Phys. Rev. D {\bf70} (2004)~125004.

\bibitem{RieffelDefQ}
M. A. Rieffel,
\textit{Deformation Quantization for Actions of $\R^d$},
Memoirs Amer. Math. Soc. {\bf 506}, Providence, RI,~1993.
 
\bibitem{Nereid}
R. Estrada, J. M. Gracia-Bond\'{\i}a and J.~C.~V\'arilly,
J. Math. Phys. {\bf30} (1989)~2789.
 
\bibitem{Melpomene}
V. Gayral, J. M. Gracia-Bond\'{\i}a and F. Ruiz Ruiz,
Nucl. Phys. B {\bf727} (2005)~513.

\bibitem{Formulas}
J.~C.~V\'arilly, E. de Far\'{\i}a and J. M. Gracia-Bond\'{\i}a,
Cienc. Tec. (Costa Rica) {\bf10} (1986)~81.

\bibitem{SantaCompanya}
L. Alvarez-Gaum\'e, F. Meyer and M. A. V\'azquez-Mozo,
``Comments on noncommutative gravity'', 
hep-th/0605113.

\bibitem{Selene}
J. M. Gracia-Bond\'{\i}a, F. Lizzi, G. Marmo and P. Vitale,
J. High Energy Phys. {\bf 0204} (2002)~026.

\bibitem{LizzSZZ...}
F. Lizzi, R. G. Szabo and A. Zampini,
J. High Energy Phys. {\bf 0108} (2001)~032.

\bibitem{GKingJohn}
S. Gutt and J. Rawnsley,
J. Geom. Phys. {\bf29} (1999)~347.

\bibitem{Vega}
J. M. Gracia-Bond\'{\i}a, F. Lizzi, F. Ruiz Ruiz and P. Vitale,
``Noncommutative spacetime symmetries: twist versus covariance'',
hep-th/0604206.

\bibitem{Overwrought}
S. Doplicher, K. Fredenhagen and J. E. Roberts,
Commun. Math. Phys. {\bf 172} (1995)~187.

\bibitem{MarcWise}
M. A. Rieffel, in \textit{Operator algebras and quantum field
theory}, International Press, Cambridge, MA, 1997; pp.~374--382.


\end{thebibliography}
\end{document}